\documentclass[aps,prd,twocolumn,showpacs,nofootinbib]{revtex4}
\usepackage{latexsym}
\usepackage{amsmath,amsfonts}
\usepackage{amsbsy}
\usepackage{mathrsfs}
\usepackage{color}
\usepackage{psfrag}
\usepackage{enumerate}
\usepackage{amsmath,amssymb,calc,amsfonts}
\usepackage{latexsym}
\usepackage{graphicx,calc,epsfig}
\def\ut#1{\rlap{\lower1ex\hbox{$\sim$}}#1{}}
\DeclareFontFamily{U}{rsfs}{}         % Formal Script            %
\DeclareFontShape{U}{rsfs}{m}{n}{<5> rsfs5 <6><7> rsfs7          %
  <8><9><10><10.95><12><14.4><17.28><20.74><24.88> rsfs10}{}     %
\DeclareMathAlphabet{\mathfs}{U}{rsfs}{m}{n}                     %
\newcommand{\mfs}[1]{\mathfs {#1}}                               %
\def\beq{\begin{eqnarray}}
\def\eeq{\end{eqnarray}}
\def\nn{\nonumber\\}

\def\del{\partial}

\def\lie{\mfs L}

\def\half{{\textstyle{\frac{1}{2}}}}

\begin{document}

\title{Hawking radiation from dynamical horizons}
\date{\today}
\author{Ayan Chatterjee}\email{ayan@tifr.res.in}
\affiliation{Tata Institute of Fundamental Research, Homi Bhabha Road, Mumbai-400005, India.}
\author{Bhramar Chatterjee}\email{bhramar.chatterjee@saha.ac.in}
\author{Amit Ghosh}\email{amit.ghosh@saha.ac.in}
\affiliation{Saha Institute of Nuclear Physics, 1/AF Bidhan Nagar, Kolkata 700064, INDIA.}

\begin{abstract}
In completely local settings, we establish that a dynamically evolving black hole horizon can be assigned a Hawking temperature. Moreover, we calculate
the Hawking flux and show that the radius of the horizon shrinks.
\end{abstract}

%\pacs{}

\maketitle

The laws of black hole mechanics in general relativity are remarkably analogous to the laws of thermodynamics \cite{Bardeen:1973gs}. This analogy is exact
 when quantum effects are taken into account. Indeed, Hawking's semiclassical analysis
establishes that quantum mechanically, a stationary black hole with surface gravity $\kappa$ radiates particles to infinity 
with a perfect black body spectrum at temperature
$\kappa/2\pi$ \cite{Hawking:1974sw}. Consequently, asymptotic observers perceive a thermal state and assign
a physical temperature to the black hole. 
The precise match to thermodynamics is complete when the thermodynamic entropy of the black hole is identified with a quarter of its area \cite{Bekenstein:1973ur}.

The original calculation of Hawking is independent of the gravitational field equations. It
relies only on the behavior of quantum fields in a specific spacetime geometry describing a stationary black hole formed due to a gravitational collapse. 
Over the years, several other techniques have been developed to study
spontaneous particle emission and the Hawking temperature for more general spacetimes. For example, the Hartle-Hawking proposal \cite{Hartle:1976tp} 
and the Euclidean approach \cite{Gibbons:1976pt} have been extensively used to associate thermal states to spacetimes with bifurcate Killing horizons. In fact, it has been established
that in any globally hyperbolic spacetime with bifurcate Killing horizon, there can exist a vacuum thermal state at temperature $\kappa/2\pi$ which remains invariant under 
the isometries generating the horizon \cite{Kay:1988mu}. 

Although these constructions are elegant, they are quite restrictive, inapplicable even for spacetimes with superradiance \cite{Kay:1988mu}.
These formulations also do not indicate how such a thermal state may arise as a result of some version of physical process. In addition, their existence requires knowledge of
global structure of spacetime. As a result, they do not appear very useful to study thermal properties of local horizons.  On the other hand,
the laws of black hole mechanics apply equally well to black hole horizons which can been proved using only
local geometrical properties of null surfaces, without any assumptions on the global development of the spacetime in which 
the horizon is embedded \cite{Hayward:1993wb, Hayward:1997jp, Ashtekar:2000sz, Booth:2003ji}. It has also been established that such 
horizons can be assigned an entropy proportional
to the area of the local horizon \cite{Ashtekar:1997yu, Ghosh:2011fc}. Thus, it seems to be a reasonable physical 
expectation that even with a local definition of black hole horizon
one should be able to establish the analogy to thermodynamics. More precisely, such horizons should have a temperature of $\kappa/2\pi$.

Incidentally, this question has been investigated in a semiclassical approach which treats Hawking radiation as a quantum tunneling phenomenon \cite{Parikh:2000ht}-\cite{Mitra:2008hj}. 
The method involves calculating the imaginary part of the action for
the (classically forbidden) process of s-wave emission, from inside and through the horizon (see \cite{Vanzo:2011tm} for more details).
 Using the WKB-approximation the tunnelling probability
for such a classically forbidden trajectory is calculated to be, $\Gamma = e^{-2{\rm Im} S}$
where, $S$ is the classical action of the trajectory to leading order in $\hbar$. 
This is equated to the Boltzmann factor $ e^{-{\beta}E}$ to extract the 
inverse Hawking temperature $\beta$. The main advantage of this formalism is that the calculations involve only the local geometry and hence can be applied to any local horizon.
Indeed, tunnelling method has been applied to local dynamical black hole horizons and the temperature is found to be $\kappa/2\pi$ where $\kappa$ is the \emph{dynamical} surface
gravity \cite{Criscienzo:2007db, Hayward:2008jq}. 
Still there are some problems with the method itself and some issues which have not been addressed in this
treatment of dynamical horizons. 
First, 
the approach depends heavily on the semiclassical approximation and though it is argued that this remains
valid near the horizon, 
it would be better to devise a more general formalism which does not rely on WKB-like approximations. Secondly, in calculating the 
imaginary part of the semiclassical action $S$ from the Hamilton-Jacobi equation, a singular integral appears with a pole at the horizon.
While for the static case the result is standard, for the dynamical horizon it is not clear how 
the integration is to be performed since the 
position of the horizon changes in a dynamical process. Lastly, in all these treatments of radiation from dynamical horizons the evolution of the horizon
itself is never addressed. In other words, it is not clear how the horizon loses area due to emission of a flux of radiation. The local formalisms of black hole 
horizon should be able to address these issues.

In this paper, a formalism is developed to establish two basic issues. First, that one can  
associate a temperature to local dynamical horizons without the need for any WKB-like approximation schemes. Second, that there
exists a precise relation between the radiation emitted by the horizon and area loss, i.e.,
flux of outgoing radiation through the horizon in between two partial Cauchy slices exactly equals the difference of radii of 
the sphere that foliates the horizon at those two instances.

We elucidate our arguments as follows. To calculate temperature for local dynamical horizons, we begin by considering the Kodama vector field \cite{Kodama:1979vn}. 
For dynamical spacetimes, this vector 
field provides a preferred timelike direction and is parallel to the Killing vector
at spatial infinity which we assume to be flat.  We can construct well-behaved positive frequency field modes on both sides of the horizon by
considering the Kodama vector field but the outgoing modes exhibit logarithmic singularities on the horizon under some approximation. However, if considered as distribution valued, these modes can be interpreted as horizon crossing and the probability current for these 
modes remain well defined. The Hawking temperature is determined if one equates the conditional probability, that modes incident
on one side is emitted to the other side, to the Boltzmann factor \cite{Damour:1976jd,Chatterjee:2012jh},
\beq
P_{(emission|incident)} = \frac{P_{(emission{\cap}incident)}}{P_{(incident)}} = e^{-{\beta}E}.
\eeq
Since this method does not depend on the entire evolution of the field modes in the spacetime, it is ideally suited for our purpose. 
    
To evaluate the Hawking flux, we recall that there are two well known (and related) definitions of local black hole horizon, the future 
outward trapping horizon (FOTH) \cite{Hayward:1993wb, Hayward:1997jp}
and the Dynamical Horizon \cite{Ashtekar:2002ag, Ashtekar:2004cn} (or its equilibrium version called the isolated horizon). 
In these local settings, black hole horizons are a stack of apparent horizons which, under suitable energy conditions,
are either null or spacelike. As such, energy flux can only remain on the surface or flow into the horizon. In order that matter 
fields flow out off such a surface requires that the surface must be timelike in some affine interval. However, to achieve a
timelike evolution of the horizon, some energy conditions need to be violated. 
This is only natural since Hawking radiation necessarily associates, with the thermal emission
of particles, a positive flux of energy flowing to infinity (we shall assume that the spacetime is asymptotically flat) and a
corresponding flux of negative energy flowing into the black hole (this negative energy flux can also be motivated by the fact that the expectation values of  
stress energy tensor of quantum fields generically violate energy conditions). In this process the horizon looses area and energy.

The plan of the paper is as follows: First, we will discuss the geometrical setup which is based on future outer trapping horizon (FOTH). 
Next, we show that how the Hawking temperature is proportional to the dynamical surface gravity associated with the Kodama vector. Finally, we will calculate the flux of energy 
radiated in a dynamical process.

We begin with definitions. We follow the conventions of \cite{Hayward:1997jp}. Consider a four dimensional spacetime $\mfs M$ with signature $(-, +,+,+)$. 
A three-dimensional submanifold $\Delta$ in $\mfs M$ is said 
to be a {\em future outer trapping horizon} (FOTH) if {\tt 1)} It is foliated by a preferred family of topological 
two-spheres such that, on each leaf $S$, the expansion $\theta_+$ of a null normal $l^{a}_+$ vanishes and the 
expansion $\theta_-$ of the other null normal $l^{a}_-$ is negative definite, {\tt 2)} The directional 
derivative of $\theta_+$ along the null normal $l^{a}_-$ (i.e., $\lie_{l_-}\theta_+$) is negative definite.

Thus, $\Delta$ is foliated by marginally trapped two-spheres. According to a theorem due to Hawking, the topology of $S$ is necessarily 
spherical in order that matter or gravitational flux across $\Delta$ is non-zero. If these fluxes are 
identically zero then $\Delta$ becomes a Killing or isolated horizon.

Even though our arguments will remain local, for definiteness, we choose a spherically symmetric background metric
\beq ds^2=-2e^{-f}dx^+dx^-+r^2(d\theta^2+\sin^2\!\theta d\phi^2)\label{spt}\eeq
where both $f$ and $r$ are smooth functions of $x^\pm$. The expansions of the two null normals are $\theta_\pm=(2/r)\,\del_\pm
r$ respectively where $\del_\pm=\partial/\del x^{\pm}$. In this coordinate system, the second requirement for FOTH translates to $\del_-\theta_+<0$
on $\Delta$.

Let the vector field $t^{a}=l^{a}_+ + h\,l^{a}_-$ be tangential to the FOTH for
some smooth function $h$. Then the
Raychaudhuri equation for $l^{a}_+$ and the Einstein equation implies
\beq \del_+\theta_+=-h\del_-\theta_+=-8\pi\, T_{++}.\label{rceq}\eeq
where $T_{++}=T_{ab}\,l^{a}_+ l^{b}_+$ and $T_{ab}$ is the energy momentum tensor. Several consequences follow from this equation. First, the 
FOTH is degenerate (or null) if and only if $T_{++}=0$ on $\Delta$. In that case, the FOTH is generated by $l^{a}_+$.
% : $T_{++}=0$
% implies $h=0$, which implies $t^2=0$. Conversely, $t^2=0$ implies $h=0$, i.e.,
% $T_{++}=0$. $\Box$
Degenerate FOTH is not interesting for Hawking radiation because this implies $\del_+r=0$. As a consequence, the 
area, $A=4\pi r^2$ of $S$, and the Misner-Sharp energy for this spacetime, given by $E=\half r$, also remains unchanged.
Secondly, since $t^2=-2h\,e^{-f}$, a FOTH becomes spacelike if and only if $T_{++}>0$ and is timelike if and only if
$T_{++}<0$.

% For a timelike FOTH, $\theta_+=0$ implies $\del_+r=0$ and $\theta_-<0$ implies
% $\del_-r<0$. As a result, 
For a timelike FOTH, several consequences follow. Here, $\lie_tr <0$, and hence,
$\Delta$ is timelike if and only if the area $A$ and the Misner-Sharp energy $E$ decreases along the horizon. This is
also expected on general grounds since the horizon receives an incoming flux of
negative energy, $T_{++}<0$. 
% This follows since $\lie_tr<0$ implies $h\del_-r<0$ and because $\del_-r<0$, it implies $h>0$,
% which implies $t$ is timelike. The converse has already been proven above.
% 
%  Proof: Follows clearly from $E=\half r$. $\Box$

As we have emphasized before, in the dynamical spacetime (\ref{spt}) the Kodama vector field plays the analog
role of the Killing vector. For this spacetime, it is given by
\beq K^{a}=e^f\,(\del_-r) \,\del^a_+-e^f\,(\del_+r)\,\del^a_-.\label{kodama}\eeq
The surface gravity is defined through $K^a\nabla_{[b} K_{a]}=\kappa\, K_b$ and is $k=-e^f\,\del_-\del_+r$. The FOTH condition
$\del_-\theta_+<0$ implies $k>0$.

Let us now determine the positive frequency modes of the Kodama vector. It is easy to see that any smooth function of $r$ is a zero-mode of the Kodama
vector.
% Proof: For any smooth function $F(r)$, $\del_\pm F=\del_\pm rF'(r)$
% and the result follows. $\Box$
Once, a zero-mode is obtained, other positive frequency eigenmodes are evaluated using 
\beq
iK\,Z_\omega=\omega \,Z_\omega \label{modeeq}
\eeq
%.
Here, $Z_\omega$ are the eigenfunctions corresponding to the positive frequency $\omega$. For simplification, let us introduce new
coordinates, $y=x^-$ and $r$ and two new 
functions, $\bar Z_\omega(y,r)=Z_\omega(x^+,x^-)$ and $G(y,r)=e^f\,(\del_+r)$.  
% Then
%
% \begin{align} &\del_+=(\del_+r)\,\del_r\\\nonumber
% &\del_-=\del_y+(\del_-r)\,\del_r.\end{align}
%
%  With this coordinate transformation, we introduce two new functions, $\bar Z_\omega(y,r)=Z_\omega(x^+,x^-)$ and $e^f\,(\del_+r)=G(y,r)$.
As a result, the eigenvalue equation \eqref{modeeq} reduces to
\beq G\,\del_y\bar Z_\omega=i\omega\,\bar Z_\omega.\eeq
Integrating and transforming 
back to old coordinates, the above equation gives
\beq Z_\omega=F(r)\exp\Big(i\omega\int_r\frac{dx^-}{e^f\del_+r}\Big)\label{modeeqnew}\eeq
where $F(r)$ is an arbitrary smooth function of $r$ and the subscript $r$ under the integral sign denotes that while doing the 
integration $r$ is kept fixed. To evaluate the integral 
in \eqref{modeeqnew}, we multiply 
the numerator and the denominator by $(\del_-\theta_+)$ and use the fact that for any fixed $r$ surface, $e^f\,(\del_-\theta_+)=-2k/r$, (although the strict interpretation of $k$ as the surface gravity holds only for surfaces with $\theta_+ =0$, it exists as a function in any neighbourhood of the horizon). Thus, in some neighbourhood of the horizon we get
\beq
\int_r\frac{dx^-\,\del_-\theta_+}{e^f\,\del_+r \del_-\theta_+}=-\int_r\frac{d\theta_+}{k
\theta_+}\eeq
We now assume (this is the only assumption we make in this calculation) that during the dynamical evolution $k$ is a slowly 
varying function in some small neighbourhood of the horizon (the zeroth law takes care of it on the horizon, but we also assume it to hold in a small neighbourhood of the horizon). This gives
\beq Z_\omega=F(r)\begin{cases} \theta_+^{-\frac{i\omega}{k}} &
\text{for}\;\theta_+>0\\
(-|\theta_+|)^{-\frac{i\omega}{k}} & {\rm for}\;
\theta_+<0.\end{cases}\label{umode}\eeq
where the spheres are not trapped `outside the trapping horizon' ($\theta_+>0$)
and fully trapped `inside' ($\theta_+<0$). These are precisely the modes which are defined outside and inside the dynamical horizon respectively
but not on the horizon. Now we have to keep in mind the modes (\ref{umode}) are not ordinary functions, but are
distribution-valued. Comparing with the spherically symmetric static case \cite{Chatterjee:2012jh}, we find for
$\epsilon\to 0^+$
\beq (\theta_++i\epsilon)^\lambda=\begin{cases} \theta_+^\lambda &
\text{for}\;\theta_+>0\\
|\theta_+|^\lambda e^{i\lambda\pi} &
\text{for}\;\theta_+<0\end{cases}\label{plusdist}\eeq
for the choice $\lambda=-i\omega/k$. The distribution (\ref{plusdist}) is
well-defined for all values of $\theta_+$ and $\lambda$, and it is
differentiable to all orders. The
modes
$Z^*_\omega$ are given by the complex conjugate distribution.
%
%\beq (\theta_+-i\epsilon)^{\lambda^*}=\begin{cases} \theta_+^{\lambda^*} &
%\text{for}\;\theta_+>0\\
%|\theta_+|^{\lambda^*} e^{-i\lambda^*\pi} &
%\text{for}\;\theta_+<0\end{cases}\label{minusdist}\eeq

We wish to calculate the probability density in a single particle Hilbert space
for positive frequency solutions across the dynamical horizon
\beq \varrho(\omega)= -\frac{i}{2}\Big[Z^*_\omega KZ_\omega-K
Z^*_\omega Z_\omega\Big]=\omega Z_\omega^*Z_\omega.\label{numberd}\eeq
A straightforward calculation gives, apart from a positive function of $r$,
\begin{align}
\varrho(\omega)&=\omega(\theta_++i\epsilon)^{-\frac{i\omega}{k}}
(\theta_+-i\epsilon)^ {\frac{i\omega}{k}}.\nn
&=\begin{cases} \omega & \text{for}\;\theta_+>0\\
\omega e^{\frac{2\pi\omega}{k}} & \text{for}\;\theta_+<0.\end{cases}
\end{align}
The conditional probability that a particle emits when it is incident on the horizon from inside is, 
\beq
P_{(emission|incident)} =   e^{-\frac{2\pi\omega}{k}}
\eeq
This gives the correct Boltzmann weight with the temperature $k/2\pi$, which is the desired value.

% Had we chosen our new coordinates to be $y=x^+$ and $r$, the Kodama vector would have taken the form $K=e^f\del_-r\del_y$ and , as a result, the eigenmodes would have been
% %
% \beq Z_\omega=F(r)\exp\Big(-i\omega\int\frac{dx^+}{e^f\del_-r}\Big).\eeq
% %
% Since, close to the horizon $e^f\del_+\theta_-=-2k/r$, we can express $Z_\omega$ in the form
% %
% \beq Z_\omega=F(r)\exp\Big(i\omega\int\frac{d\theta_-}{k\theta_-}\Big).\eeq
% %
% So under the same approximation of slowly varying $k$, $Z_\omega=F(r)(\theta_-+i\epsilon)^{i\omega/k}$.

We now show that as the horizon evolves, the radius of the 2-sphere foliating the horizon shrinks in precise accordance with the 
amount of flux radiated by the horizon. To study the flux equation, consider new coordinates, ($x^+,x^-)\mapsto
(\theta_+,\tilde x^-)$ where $\tilde x^-=x^-$. 
% As a result,
%
% \begin{align} &dx^-=d\tilde x^-\\ &dx^+=\left(\frac{1}{\del_+\theta_+}\right)\, d\theta_+-
% \left(\frac{\del_-\theta_+}{\del_+\theta_+}\right) \,d\tilde x^-.\end{align}
%
On FOTH, $(\del_-\theta_+)/(\del_+\theta_+)$ is equal to
$-(\del_-\del_+r)/(4\pi r\,T_{++})$ and negative definite. As a result, the
derivatives are related to each other by
\beq \tilde\del_-=\del_-+\left(\frac{\del_-\del_+r}{4\pi
r\,T_{++}}\right)\,\del_+.\label{tilded}\eeq
It is not difficult to show that $\tilde\del_-$ is proportional to the tangent vector
$t^a$ to the FOTH. 
Observe that
the normal one-form to $\Delta$ must be proportional to
$(dr-2\,dE)$, which on the horizon is equal to the one-form
\beq (8\pi e^fr^2\,T_{++}-2re^f\del_-\del_+r)\,\del_-r \, dx^-.\eeq
In arriving at the above identity we have made use of two Einstein's
equations \cite{Hayward:1997jp}
\begin{eqnarray}\label{Einsteineqn}
 r\,\del_-\del_+r+\del_+r\,\del_-r+\half e^{-f}&=&4\pi r^2\,T_{-+},\\ \nonumber
\del_+^2r+\del_+f\,\del_+r&=&-4\pi r\,T_{++},
\end{eqnarray}
%
 %$r\del_-\del_+r+\del_+r\del_-r+\half e^{-f}=4\pi r^2T_{-+}$ and 
%$\del_+^2r+\del_+f\del_+r=-4\pi rT_{++}$ along with the
and energy equations
\beq \del_\pm E=2\pi e^f r^3(T_{-+}\,\theta_\pm-T_{\pm\pm}\,\theta_{\mp}).\eeq
As a result, the normal vector $n^a$ is proportional to
\beq \del_+-\left(\frac{4\pi
r\,T_{++}}{\del_-\del_+r}\right)\del_-=\del_+-h\del_-,\eeq
so that the tangent vector $t^a=\del^a_++h\del^a_-$, which is clearly proportional to
(\ref{tilded}). 

%$\Box$

So $\tilde x^-,\theta,\phi$ are natural coordinates on FOTH. The
line-element (\ref{spt}) induces a line-element on $\Delta$
\beq ds^2=-2e^{-f}h^{-1}(d\tilde x^-)^2+r^2(d\theta^2+\sin^2\!\theta
d\phi^2).\eeq
Consequently, the volume element on $\Delta$ is given by $d\mu=\sqrt{2e^{-f}h^{-1}}r^2\sin\theta\,d\tilde x^-d\theta d\phi$. We can
now calculate the flux of matter energy that crosses the dynamical horizon---it is an integral on a slice of horizon bounded 
by two spherical sections $S_1$ and $S_2$
\beq \mfs F=\int d\mu\;T_{ab}\hat n^a K^b\eeq
where $\hat n^a$ is the unit normal vector 
\beq \hat n^a=\frac{1}{\sqrt{2he^{-f}}}(\del_+^a-h\del_-^a)\eeq
and $K^a$ is the Kodama vector. Using spherical symmetry, eqn. (\ref{tilded}) and eqn. \eqref{Einsteineqn}, we get

%  a simple calculation gives
%
 \begin{align} \mfs F&=\int d\tilde x^-\;4\pi r^2(\frac{1}{h}T_{++}-T_{+-})e^f\del_-r\nn
 &=\int d\tilde x^-\;4\pi r^2(\frac{1}{4\pi r}\del_+\del_-r-T_{+-})e^f\del_-r.
 \end{align}\label{flux1}
Making use of the Einstein equation (\ref{Einsteineqn}) on the horizon 
and (\ref{tilded}), we get
\begin{align} \mfs F&=-\int d\tilde x^-\;\frac{1}{2}\del_-r=-\int d\tilde x^-\;\frac{1}{2}\tilde\del_-r\nn &=-\frac{1}{2}(r_2-r_1)\label{flux} \end{align}
where $r_1,r_2$ are respectively the two radii of $S_1,S_2$. Since the area is decreasing along the
horizon, $r_2<r_1$ where $S_2$ lies in the future of $S_1$. As a result, the outgoing flux of matter energy 
radiated by the dynamical horizon is positive definite (and the ingoing flux of matter energy is negative definite).
The flux formula (\ref{flux}) differs from that given in \cite{Ashtekar:2002ag}. Since the Kodama vector field provides a timelike direction
and is null on the horizon, it seems more appropriate to use $K^a$ for the dynamical horizon. 
  
The derivation of Hawking temperature and the flux law depends on two assumptions. First, that the Kodama vector exists in the spacetime. For spherically symmetric 
spacetimes, the Kodama vector field exists unambiguously and the Misner-Sharp energy is well defined. For more general spacetimes,
a Kodama-like vector field is not known, however, one can still define some mass for such cases that reduces to the Misner-Sharp energy in the 
spherical limit \cite{Mukohyama:1999sp}. The second
assumption, the existence of a slowly varying $k$ can also be motivated for large black holes. In such cases, the horizon evolves slowly enough so that the surface gravity function should vary slowly in some small neighbourhood
of the horizon. Alternatively, we can conclude that the Hawking temperature for a dynamically evolving large black hole is $k/2\pi$ if the dynamical surface gravity  is slowly varying in the vicinity of the horizon. 

The set-up described in this paper can be further developed to model dynamically evaporating black hole horizons through Hawking radiation, analytically as well as numerically. Over 
the years, several models have been constructed 
which study radiating black holes, formed in a gravitational collapse, based on the imploding Vaidya metric with a negative energy-momentum tensor, show that a timelike apparent horizon
forms due to violation of energy conditions \cite{Hiscock:1980ze}.
However, such models are based on global considerations of event horizons, while local structures like that used in \cite{Hayward:2005gi}
might be useful for a better understanding of Hawking radiation and computations of quantum field theoretic effects (see also \cite{Pranzetti:2012pd}).

It is also interesting to speculate on the extension of the present method for other diffeomorphism invariant theories of gravity. While the zeroth and the first law hold for 
any arbitrary such theory, the second law has only been proved for a class of such theories \cite{Chatterjee:2011wj}. If the present formalism can be 
extended to other theories of gravity, it will lend a support to the existence of the area increase theorem for such theories.

While more interesting and deeper issues can only be understood in a full quantum theory of gravity, the present framework can elucidate the suggestions of \cite{Ashtekar:2005cj}
and provide a better understanding of the Hawking radiation process.

\end{document}